\documentclass[twocolumn]{aastex631}

\usepackage{silence}
\DeclareRobustCommand{\ion}[2]{\textup{#1\,\textsc{\lowercase{#2}}}}
\usepackage{graphicx}
\usepackage{savesym}
\savesymbol{tablenum}
\usepackage{siunitx}
\restoresymbol{SIX}{tablenum}
\usepackage{natbib}
\newcommand{\doitoggle}[1]{}

\usepackage{amsmath}
\usepackage{txfonts}
\usepackage{multirow}
\usepackage{float}

\usepackage{comment}
\usepackage{ragged2e}
\usepackage{xcolor}
\begin{document}

\title{Mapping a Quasar Outflow from Parsec to Kiloparsec Scales: \\ A Combined HST Absorption and VLT Emission Investigation}

\author[0009-0001-5990-5790]{Mayank Sharma}
\affiliation{Department of Physics, Virginia Tech, Blacksburg, VA 24061, USA}
\email{Corresponding Author: mayanksh@vt.edu}

\author[0000-0002-3365-8875]{Travis C. Fischer}
\affiliation{AURA for ESA, Space Telescope Science Institute, 3700 San Martin Drive, Baltimore, MD 21218, USA}

\author[0000-0002-4314-021X]{Manuela Bischetti}
\affiliation{Dipartimento di Fisica "Enrico Fermi", Università di Pisa, Largo Bruno Pontecorvo 3, Pisa, I-56127, Italy}
\affiliation{INAF - Osservatorio Astronomico di Trieste, Via G. B. Tiepolo 11, I–34131 Trieste, Italy}

\author[0000-0003-2991-4618]{Nahum Arav}
\affiliation{Department of Physics, Virginia Tech, Blacksburg, VA 24061, USA}

\author[0000-0002-4031-4157]{Fabrizio Fiore}
\affiliation{INAF - Osservatorio Astronomico di Trieste, Via G. B. Tiepolo 11, I–34131 Trieste, Italy}
\affiliation{Dipartimento di Fisica “G. Occhialini”, Università degli Studi di
Milano-Bicocca, Piazza della Scienza 3, 20126 Milano, Italy}

\author[0000-0002-4227-6035]{Chiara Feruglio}
\affiliation{INAF - Osservatorio Astronomico di Trieste, Via G. B. Tiepolo 11, I–34131 Trieste, Italy}

\author[0009-0009-7058-5415]{Manoj Ghosh}
\affiliation{Department of Physics, Virginia Tech, Blacksburg, VA 24061, USA}

%% Note that the \and command from previous versions of AASTeX is now
%% depreciated in this version as it is no longer necessary. AASTeX 
%% automatically takes care of all commas and "and"s between authors names.

%% AASTeX 6.31 has the new \collaboration and \nocollaboration commands to
%% provide the collaboration status of a group of authors. These commands 
%% can be used either before or after the list of corresponding authors. The
%% argument for \collaboration is the collaboration identifier. Authors are
%% encouraged to surround collaboration identifiers with ()s. The 
%% \nocollaboration command takes no argument and exists to indicate that
%% the nearby authors are not part of surrounding collaborations.

%% Mark off the abstract in the ``abstract'' environment. 
\begin{abstract}
Linking nuclear winds to galactic-scale outflows remains a major observational challenge in understanding the multiscale physics of active galactic nuclei feedback. Here we present VLT/KMOS integral-field spectroscopy and SDSS observations of the $z = 0.9655$ quasar PKS J0352$-$0711. Our analysis reveals complex, multi-ionization emission, including a fast, unresolved nuclear wind and a spatially resolved galactic-scale outflow. We integrate the [\ion{O}{III}] emission properties with those deduced from the mini-broad-absorption-line outflows detected in HST/COS observations of this quasar. This unique combination of datasets allows us to trace, for the first time, the physical progression of a quasar outflow from $\sim$ 10 pc to 10 kpc. The multiscale kinematics support a unified evolutionary scenario where the inner, constant-velocity ($\sim-3800 \textrm{ km s}^{-1}$) expansion of the wind is traced jointly in absorption ($\sim 9$ pc) and emission ($\gtrsim 40$ pc). As the wind propagates to $\sim$ 500 pc, the intermediate absorption system reveals a deceleration to $\sim-2100 \textrm{ km s}^{-1}$, consistent with mass-loading from the interstellar medium. Finally, our spatially resolved observations capture the gas breaking out of the inner galaxy, in the form of a wide-angle blueshifted outflow expanding beyond 8 kpc, with a velocity of $\sim -1000 \textrm{ km s}^{-1}$. Despite the three orders of magnitude variation in spatial scale, and a factor-of-four deceleration, the momentum fluxes remain consistent within uncertainties across all scales. These results suggest that the distinct outflow components represent the integrated history of a sustained feedback cycle from nuclear to galactic scales. 
\end{abstract}
%% Keywords should appear after the \end{abstract} command. 
%% The AAS Journals now uses Unified Astronomy Thesaurus concepts:
%% https://astrothesaurus.org
%% You will be asked to selected these concepts during the submission process
%% but this old "keyword" functionality is maintained in case authors want
%% to include these concepts in their preprints.
\keywords{Broad-absorption line quasar (183), Active galactic nuclei (16), AGN host galaxies (2017), Interstellar medium (847), Galactic winds (572), Galaxy evolution (594)}

%Galaxies (573), Active galactic nuclei (16), Quasars (1319), AGN host galaxies (2017)}

%% From the front matter, we move on to the body of the paper.
%% Sections are demarcated by \section and \subsection, respectively.
%% Observe the use of the LaTeX \label
%% command after the \subsection to give a symbolic KEY to the
%% subsection for cross-referencing in a \ref command.
%% You can use LaTeX's \ref and \label commands to keep track of
%% cross-references to sections, equations, tables, and figures.
%% That way, if you change the order of any elements, LaTeX will
%% automatically renumber them.
%%
%% We recommend that authors also use the natbib \citep
%% and \citet commands to identify citations.  The citations are
%% tied to the reference list via symbolic KEYs. The KEY corresponds
%% to the KEY in the \bibitem in the reference list below. 

\section{Introduction}
Active galactic nuclei (AGN) feedback is vital for our understanding of the coevolution of supermassive black holes (SMBHs) and their host galaxies. Although the active phase of SMBHs is relatively short-lived \citep[$\sim10^7$ yrs, e.g.,][]{hopkins2006evolution, lin2022agn}, the resulting injection of momentum and energy effectively regulates the long-term evolution of the host galaxies and their environment \citep[e.g.,][]{silk1998quasars,fabian2012observational}. Cosmological simulations and semi-analytical models of galaxy formation have invoked AGN feedback to explain: the tight relationship between the SMBH mass and the bulge-properties \citep[e.g.,][]{di2005energy,marsden2020case,shankar2025probing}, the cooling flow problem in X-ray clusters \citep[e.g.,][]{peterson2006x,mcnamara2007heating,lehle2024heart,chen2026cool} and the high-end cutoffs for the mass and luminosity functions of galaxies \citep[e.g.,][]{croton2006many,bower2006breaking,schaye2015eagle,ni2022astrid}. \par
For luminous quasars accreting at high rates, outflows driven by the central engine are considered the primary agents of feedback \citep[e.g.,][]{silk1998quasars,hopkins2010quasar,king2015powerful}. These outflows manifest across a diverse range of spatial scales and ionization states. At nuclear scales, they are identified primarily through blueshifted absorption features. In the innermost sub-parsec regions, these appear as highly ionized Ultra-Fast Outflows (UFOs) detected in X-rays, with typical velocities of $\sim0.1-0.3c$ \citep[e.g., in quasars PG 1211 and PDS 456;][]{danehkar2018ultra, 2025Natur.641.1132X}. Further out, their primary signature takes the form of UV absorption-line outflows. Based on the velocity widths, they are classified as either broad absorption line (BAL, $\Delta \textrm{v} \gtrsim$ 2000 km s$^{-1}$), mini-broad absorption line (mini-BAL, $500 \lesssim \Delta \textrm{v} \lesssim$ 2000 km s$^{-1}$) or narrow absorption line (NAL, $\Delta \textrm{v} \lesssim$ 500 km s$^{-1}$) outflows. In these different forms, they span a large range of velocities ($\sim100-50,000$ km s$^{-1}$) and distances from sub-pc to several kpc \citep[e.g.,][]{moe_quasar_2009,arav2018evidence,arav2020hst,xu2019vlt,miller2020hstb,rodriguez2020survey,choi_physical_2022,dehghanian_quasar_2025,sharma_desi_2025}. While these inner components are seen in absorption, integral field spectroscopy (IFS) and interferometric observations spatially resolve emission lines in ionized, molecular and atomic phases. These emission signatures have revealed outflowing structures that extend from the inner interstellar medium (ISM) out to the circumgalactic medium \citep[e.g.,][]{feruglio2010quasar,liu2013observations,fischer2019spatially,bischetti2019gentle,bischetti2025alma,veilleux2023first,zhao2023discovery}.\par  
To better understand the mechanics of AGN feedback, it is essential to infer the relationship between these diverse outflows across multiple phases and spatial scales. Establishing this connection allows us to trace the outward propagation of nuclear winds as well as their interaction with the host galaxy's ISM. The sample of sources in which multiphase and multiscale outflows have been studied simultaneously remains severely limited \citep[see e.g. the compilation in][]{baldini2024winds}. Current efforts have primarily focused on linking the inner-most X-ray UFOs directly to large-scale molecular (e.g., CO and OH) or ionized (e.g., [\ion{O}{III}] and H$\beta$) emission outflows \citep[e.g.,][]{feruglio2017discovery,veilleux2017quasar, smith2019discovery, bellocchi2026multi}. These studies indicate that nuclear winds are energetically capable of driving the observed galactic-scale outflows. However, the vast spatial separation between these phases has left the intermediate parsec-to-hundred-parsec region unexplored. \par
UV absorption-line outflows offer a unique opportunity to trace this scale, but their connection with the large-scale outflows has been studied in an even smaller number of sources ($<$ 10). Among these studies \citet{liu2015integral}, \citet{perna2025ga} and \citet{sharma2025distance, zhao2025galactic} find that the BAL outflows in their sources trace the same spatial scale ($1-10$ kpc) as the [\ion{O}{III}] outflows. On the other hand, \citet{nardini2018multi} and \citet{bischetti2024multiphase} find evidence for BAL outflows at sub-kpc scales along with kpc scale outflows traced in emission in their sources. However, the lack of density-sensitive troughs in their spectra leads to a two-order-of-magnitude uncertainty in the derived distance and energetics. Thus, the evolution of the outflowing gas in this crucial spatial scale, where most of its interaction with the ISM is predicted to occur, has largely remained an enigma. \par
In this work, we present an IFS study of PKS J0352--0711 (hereafter PKS 0352), a luminous quasar at $z = 0.9655$, with a bolometric luminosity of $L_{\textrm{bol}}=5.5\times10^{46} \textrm{ erg s}^{-1}$. PKS 0352 was previously targeted by the Cosmic Origins Spectrograph (COS) aboard the \textit{Hubble} Space Telescope (HST). In their analysis of these data, \citet{miller2020hst} reported the detection of two mini-BAL outflows with maximum velocities of $\sim$ $-3700$ and $-2100$ km s$^{-1}$. Both systems show troughs from ions with a broad range in ionization potential (IP), ranging from \ion{N}{III} (IP $\sim 47$ eV) to \ion{Mg}{X} (IP $\sim 367$ eV). Through photoionization and excited state trough analyses, the distances of these outflows from the central source were found to be $\sim 10$ pc and $\sim 500$ pc, respectively. The discovery of these winds with multiple ionization states across nearly two orders of magnitude in spatial scale motivated a search for their galactic-scale counterparts. \par 
Here, we report the detection of extended [\ion{O}{III}] emission out to $\sim10$ kpc in PKS 0352, which reveals two spatially and kinematically distinct outflowing components. We find that the multi-ionization structure also exists in emission, as confirmed through the detection of an unresolved [\ion{Ne}{V}] outflow. By connecting these emission tracers with the absorption components, we present the physical evolution of the outflowing gas across three orders of magnitude in distance.\par
The paper is organized as follows. We first describe the observations and data reduction in Section \ref{sec:obs_datared}. In Section \ref{sec:data_ana}, we present the analysis of our multiwavelength data, characterizing the morphology and kinematics of the different outflowing components. We estimate their energetics in Section \ref{sec:energetics} and discuss the connection between the multiscale outflows in Section \ref{sec:physcon}. In Section \ref{sec:discussion}, we discuss the significance of two unique [\ion{O}{III}] features discovered in our analysis and conclude with a summary in Section \ref{sec:summary}. Throughout this paper, we adopt a $\Lambda$CDM cosmology with $H_0 = 69.6$ km s$^{-1}$ Mpc$^{-1}$, $\Omega_m = 0.286$ and $\Omega_\Lambda = 0.714$ \citep{bennett20141}. Using the \citet{wright2006cosmology} calculator, this results in a physical scale of 8.05 kpc per arc second at the redshift of our source. 

\section{Observations and Data Reduction} \label{sec:obs_datared}

PKS 0352 was observed with the K-band Multi Object Spectrograph \citep[KMOS,][]{sharples2013first} on the Very Large Telescope (VLT) in August 2025, as a part of program 0115.B-2246 (PI: Bischetti). These observations used the IZ grating, which provides a rest-frame wavelength coverage of 4000$-$5375 \r{A} with a spectral resolution of $R\sim 3400$. The targeted field of view is \ang{; ; 2.8} $\times$ \ang{; ; 2.8}, which is sampled with a spatial pixel scale of \ang{; ; 0.20} $\times$ \ang{; ; 0.20} in our seeing-limited observations. \par
We utilized the multi-IFU setup of KMOS to obtain simultaneous science and sky exposures, enabling efficient background subtraction. Over two nights, 12 exposures of 325 seconds each were obtained, for a total exposure time of 3900 seconds. For this observational setup, an  average seeing of \ang{; ; 0.63} was reported by ESO's seeing monitor. \par
The reduced dataset was made available by ESO as part of their KMOS Phase 3 data stream. The data reduction was performed using version 4.5.1 of the KMOS pipeline \citep[][]{davies2013software} and included sky subtraction, flat-fielding, wavelength calibration, flux calibration, and cube reconstruction. This resulted in two reduced data cubes, each corresponding to combined observations from a single night. These were co-added after verifying the source position in both cubes, yielding the final data cube for our analysis. \par
PKS 0352 was also observed as part of the Sloan Digital Sky Survey (SDSS). This legacy observation was obtained using the original SDSS spectrograph in December 2000, with a total exposure time of 2700 seconds. The reduced archival spectrum for this observation was retrieved from Data Release 17 \citep{accetta2022seventeenth} and provides rest-frame wavelength coverage of 1940$-$4690 \r{A}, with a spectral resolution of $R\sim 2000$. 
\begin{figure}[t]
   \centering
   \resizebox{\hsize}{!}
        {\includegraphics[width=0.80\linewidth]{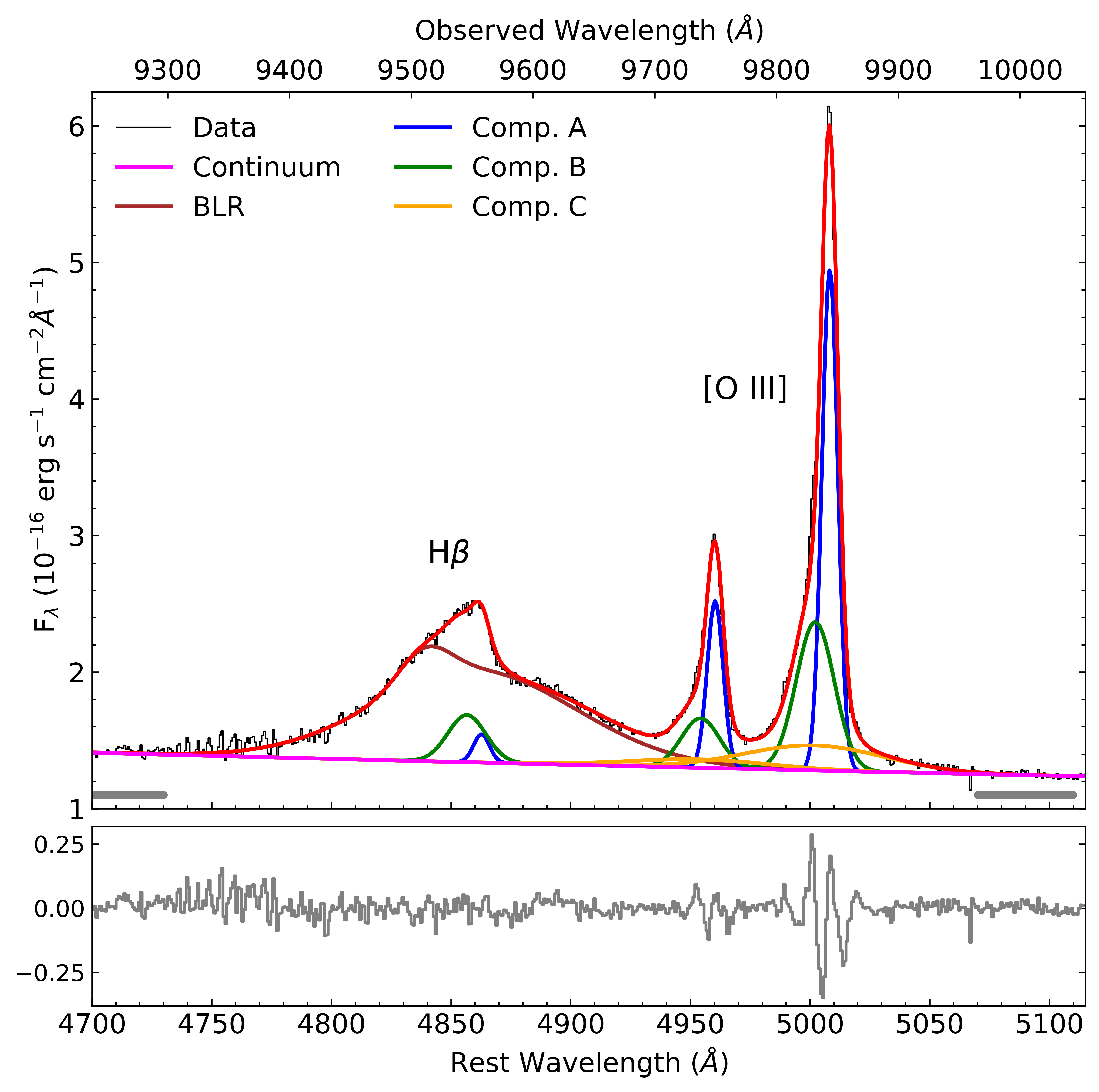}}
      \caption{VLT/KMOS spectrum of the [\ion{O}{III}] region in PKS 0352, extracted from the central 3 $\times$ 3 spaxels. The top panel shows the observed spectrum in black and the total best-fit BEAT model in red. The power-law continuum (magenta) is fit using the spectral regions marked by the gray bars. The combined H$\beta$ BLR model is shown in brown. The blue curve represents the systemic core of the NLR emission, while the green and orange curves trace the broad outflowing components. The bottom panel shows the unnormalized fit residual in gray.}
         \label{fig:central_model}
   \end{figure}
  
\section{Analysis and Results} \label{sec:data_ana}

\subsection{Unresolved Optical Emission} \label{subsec:nuclear}

We characterize the emission from the nuclear region by extracting a spectrum from the central 3 $\times$ 3 spaxels. This spectrum reveals multiple emission features from the broad line region (BLR) traced by the H$\beta$ $\lambda$4861 and H$\gamma$ $\lambda$4340 lines, as well as forbidden line emission, traced by the [\ion{O}{III}] $\lambda \lambda$ 4959, 5007 doublet. To isolate these line features, we first model the underlying continuum and \ion{Fe}{II} emission. By selecting spectral regions free from any line emission, we find that a broken power-law profile provides a good fit to the continuum. We also include the \ion{Fe}{II} pseudo-continuum template described in \citet{kovavcevic2010analysis} and \citet{Shapovalova2012} in our model, finding contamination from the \ion{Fe}{II} emission to be negligible in the KMOS spectrum of PKS 0352.  \par
After subtracting the total underlying continuum, we narrow our focus to the H$\beta$-[\ion{O}{III}] spectral region (Figure \ref{fig:central_model}) where the strongest emission-line features are detected. These lines exhibit a complex structure, as evidenced by their strong asymmetry. We model them with multi-component Gaussian profiles using the Bayesian Evidence Analysis Tool (BEAT). The BEAT algorithm was introduced in \citet{fischer2017gemini} and uses the nested sampling package \textsc{Ultranest} \citep{buchner2021ultranest} to fit models with increasing complexity (increasing number of Gaussians in our case) to the data until a more complex model is no longer preferred with a $> 99\%$ posterior probability. The statistical principle behind BEAT and its advantages over traditional reduced-$\chi^2$ methods are described in the Appendix of \citet{fischer2017gemini} and references therein. In our models, we kinematically tie together the velocity centroid ($v_0$) and dispersion ($\sigma$) of the non-BLR emission from H$\beta$ to those of the [\ion{O}{III}] $\lambda \lambda$ 4959, 5007 doublet, while the BLR emission is modeled with independent components. The fluxes of all emission lines are allowed to vary freely, except for the [\ion{O}{III}] $\lambda \lambda$ 4959, 5007 doublet, where the relative strength is fixed to the theoretical expectation of 3:1. With these constraints, we fit the spectrum with BEAT and show the resulting highest-likelihood model in Figure \ref{fig:central_model}. Based on this model, we find that the BLR emission requires a sum of two Gaussians to capture its asymmetric profile. Furthermore, the forbidden line emission, traced by the [\ion{O}{III}] $\lambda \lambda$ 4959, 5007 doublet, is decomposed into three different Gaussians, which we label alphabetically in order of increasing width. \par
\begin{figure*}[ht!]
   \centering
    \includegraphics[width=0.85\linewidth]{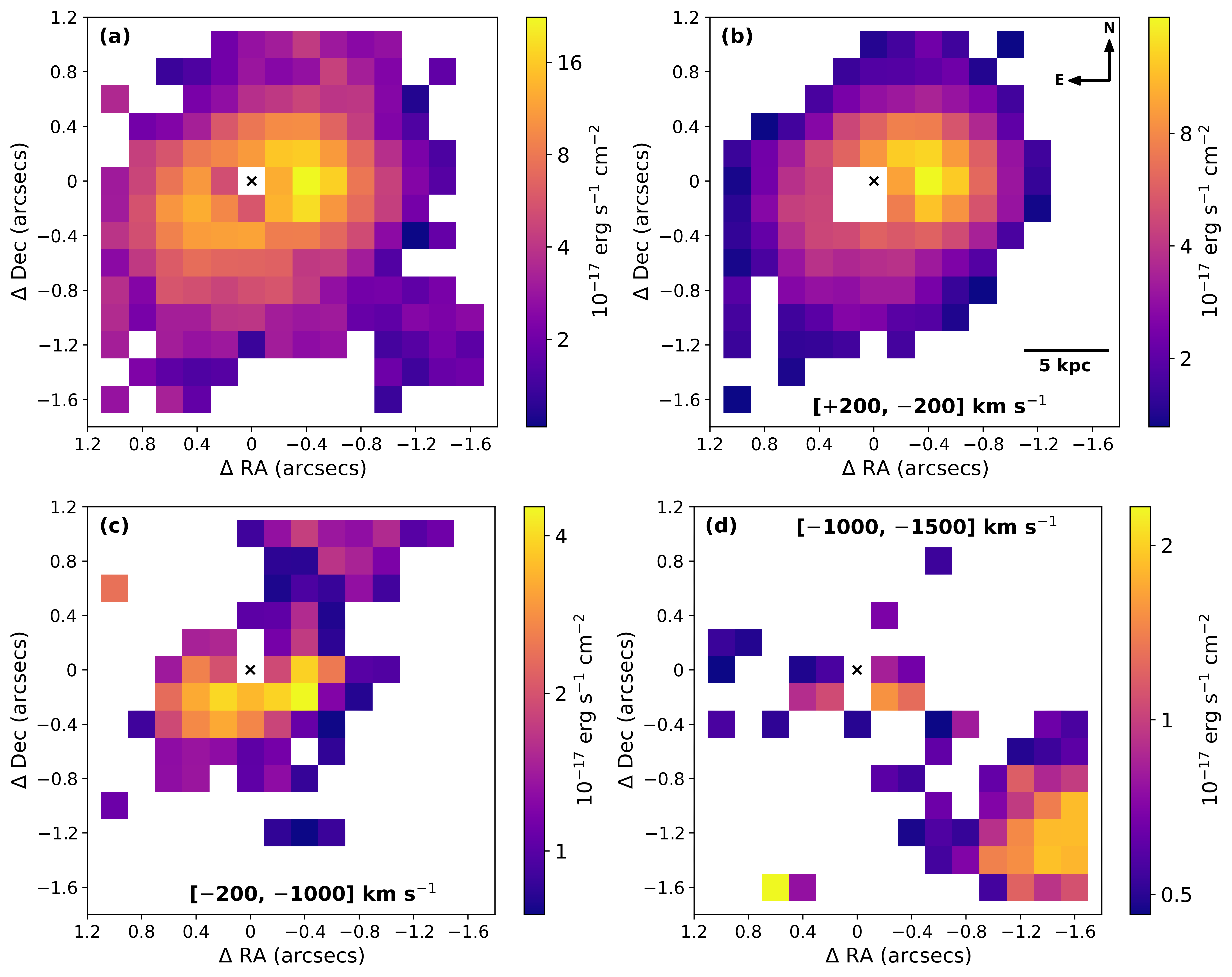}
    \caption{PSF-subtracted flux maps of the extended [\ion{O}{III}] emission around PKS 0352. (a) Total integrated flux map. (b) Emission near systemic velocities with $|v| < 200 \textrm{ km s}^{-1}$. (c) Highly blueshifted gas between $-200$ and $-1000 \textrm{ km s}^{-1}$. (d) The highest velocity blueshifted emission between $-1000$ and $-1500 \textrm{ km s}^{-1}$. The central cross marks the position of the quasar. }
         \label{fig:psfsubmaps}
   \end{figure*}

The narrowest component (Comp.\ A) has a full width at half maximum (FWHM) of 470 km s$^{-1}$. We associate this with systemic emission from the host galaxy's narrow line region (NLR) and thus use its centroid velocity to derive a redshift for PKS 0352 of $z$ = 0.9655. This is consistent within 70 km s$^{-1}$ with the value of 0.9662 obtained from the SDSS spectra by \citet[][]{hewett2010improved} (using the spectrally unresolved [\ion{O}{II}] $\lambda \lambda$ 3726, 3729 doublet) and used by \citet{miller2020hst} in their analysis. \par
Comp.\ B, on the other hand, is much broader, exhibiting a FWHM of 1120 km s$^{-1}$ and a blueshifted centroid velocity of $-$360 km s$^{-1}$. Because this width significantly exceeds the maximum expected from gravitational motion alone \citep[$\sim$ 600 km s$^{-1}$, e.g.,][]{harrison2016kmos}, we interpret Comp.\ B as a signature of outflowing gas. While Comps.\ A and B together can adequately explain the non-BLR emission seen in H$\beta$, they fail to reproduce the broad wings of the [\ion{O}{III}] doublet. This necessitates Comp.\ C, which shares a similar blueshifted velocity to Comp.\ B ($-$390 km s$^{-1}$), but a remarkably larger FWHM of 4130 km s$^{-1}$. This suggests the presence of a more extreme phase of the outflow, whose significance we discuss in Section \ref{subsec:extreme}. In Table \ref{table:nuclear_kinematics}, we summarize the kinematic properties of all three [\ion{O}{III}] components.  

\begingroup
\setlength{\tabcolsep}{4pt}
\renewcommand{\arraystretch}{1.5}
\begin{table}
\caption{Kinematic properties of the unresolved nuclear gas.}
\centering
\begin{tabular}{lcccc}
\hline
\hline
Ion &Component & $v_0$ & FWHM & ${v_{\textrm{max}}}^{a}$ \\
\hline
\multirow{3}{*}{\ion{O}{III}}&A & $0\pm8$ & $470\pm19$ & $-$ \\
%$-417\pm18$\\
&B & $-360\pm21$ & $1120\pm35$ & $-1315\pm36$\\
&C & $-390\pm77$ & $4130\pm226$ & $-3900\pm207$\\
\multirow{2}{*}{\ion{Ne}{V} $^b$}& A & $-5\pm20$ & $435\pm165$ & $-$\\
& B & $-370\pm165$ & $1330\pm120$ & $-1505\pm106$\\
\multirow{2}{*}{\ion{Ne}{III} $^b$}& A & $16\pm20$ & $300\pm150$ & $-$\\
& B & $-230\pm150$ & $1155\pm160$ & $-1215\pm140$\\
\hline
\end{tabular}
\tablecomments{a: Maximum velocity of the outflowing components, defined as $v_{\textrm{max}}$ = $v_{0}~-$ 2$\sigma$ \citep{rupke2013multiphase, fiore2017agn}, where $\sigma = \textrm{FWHM/2.355}$. b: See Section \ref{subsec:UVnuclear} and Figure \ref{fig:NeV_model}.}
\label{table:nuclear_kinematics}
\end{table}
\endgroup

\subsection{Spatially Resolved Optical Emission} \label{subsec:resolved}

\begin{figure*}[t]
   \centering
   \resizebox{\hsize}{!}
        {\includegraphics[width=1.00\linewidth]{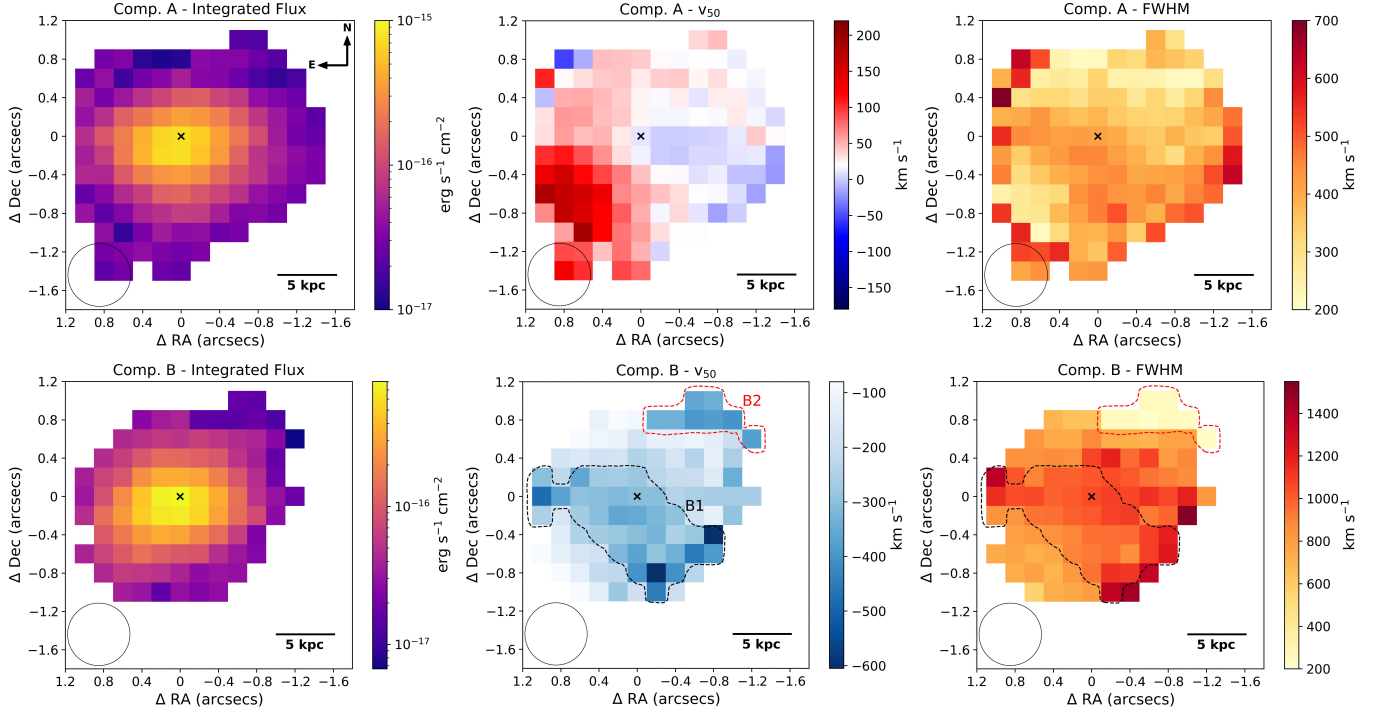}}
      \caption{Spatially resolved maps of the two distinct components of the [\ion{O}{III}] line emission, showing only spaxels with a $>2\sigma$ detection. The left panels show the integrated flux, while the middle and right panels trace the velocity centroid ($v_{50}$) and line width (FWHM), respectively. Component A (top row) traces the large-scale rotating disk of the host galaxy, while Component B (bottom row) represents the blueshifted outflowing gas. In the middle and right panels of the bottom row, black and red contours indicate regions B1 and B2, respectively, which contain the bulk of the outflowing gas with high blueshifted velocities ($v_{50}<-250 \textrm{ km s}^{-1}$). The central black cross marks the location of the quasar and the black circle in the lower-left corner of each panel represents the FWHM of the instrumental PSF.}
         \label{fig:maps}
   \end{figure*}

\subsubsection{PSF-Subtracted Channel Maps}

To reveal the presence of any spatially resolved emission around PKS 0352, we construct flux maps after removing the point source from the quasar position. Following the approach of \citet{zhao2025galactic}, we use the line-free continuum region between 5070-5150 \r{A} as a tracer of the quasar point source emission. By integrating the flux within this spectral range, we extract the point spread function (PSF) profile shown in the Appendix. The measured FWHM of $\sim$ \ang{; ; 0.65} (estimated via a circular fit) is in good agreement with the average seeing during the observations (\ang{; ; 0.63}). We scale the spectrum from the central spaxel using the empirical PSF template and subtract it from each individual spaxel. The resulting integrated [\ion{O}{III}] flux map (panel (a) of Figure \ref{fig:psfsubmaps}) reveals a complex, spatially-resolved structure extending on large scales across the field of view. \par
We probe the kinematics of this extended gas with the help of narrow-band images obtained by collapsing the PSF-subtracted cube along different velocity channels. Panel (b) of Figure \ref{fig:psfsubmaps} traces emission close to systemic velocities ($|v| <200 \textrm{ km s}^{-1}$), showing a broad, relatively symmetric spatial distribution. In contrast, panel (c) captures highly blueshifted gas (between $-200$ and $-1000 \textrm{ km s}^{-1}$) which shows a structure characterized by extended, asymmetric regions of bright emission. Finally, panel (d) isolates the highest velocity gas between $-1000$ and $-1500 \textrm{ km s}^{-1}$. This emission is centered at the southwestern edge of the field of view, appearing physically detached from the main emission structures. We defer the discussion of this isolated feature to Section \ref{sec:tidal} and focus here on the extended emission surrounding the central quasar. \par

\subsubsection{Emission Line Fitting}

To further study the dichotomy between the emission structures revealed by panels (b) and (c) of Figure \ref{fig:psfsubmaps}, we model the emission in the H$\beta$-[\ion{O}{III}] spectral region for individual spaxels. For each spaxel, we first subtract the power-law continuum and \ion{Fe}{II} emission following the same procedure as described in Section \ref{subsec:nuclear}. For this continuum-subtracted cube, we use BEAT to obtain independent fits for the line emission in every spaxel. Since the off-nuclear spaxels contain contribution from the unresolved BLR due to the instrumental PSF, we fix the shape of the BLR emission in the models and allow only its flux to vary between spaxels. \par
As with our nuclear fit, we rely on Bayesian evidence to determine the number of Gaussian components, requiring a posterior probability $>99\%$ to favor a more complex model. Additionally, to avoid fitting noise features, we also require each independent component to be detected with a  S/N $>$ 2. In our initial iteration of the fits, we find that spaxels within the central $\sim$ \ang{; ; 0.5} diameter require up to three Gaussian components. The broadest of these displays kinematics similar to the nuclear component C, with little variation between spaxels. This suggests that it arises from unresolved emission within the nuclear region that has been smeared by the PSF. Consequently, we update our models to include the nuclear template of Comp.\ C, with its flux tied to that of the BLR. In the resulting best fits, we find that the spatially resolved structure of the [\ion{O}{III}] emission is adequately modeled with up to two Gaussian components. Because multi-Gaussian decomposition can be susceptible to fitting degeneracies, we examine the parameter distributions of the best-fit centroid velocities and line widths of the two components across all modeled spaxels, following the approach of \citet{vayner2024first}. We find a bimodal structure for both distributions where one component is confined to low velocities and narrow widths, whereas the other primarily traces high-velocity, high dispersion gas. This strongly suggests that the two components trace distinct physical structures, and we thus use the best-fit parameters to reveal their spatially resolved morphology and kinematics, as shown in Figure \ref{fig:maps}. \par 
Component A (top row, Figure \ref{fig:maps}) cleanly isolates the relatively symmetrical spatial distribution initially observed in our low-velocity channel map (panel (b) of Figure \ref{fig:psfsubmaps}). It is characterized by low centroid velocities ($|v_{50}|<200 \textrm{ km s}^{-1}$) and narrow widths (FWHM $\lesssim$ 600 km s$^{-1}$), properties consistent with the gas intrinsic to the host galaxy. This emission is slightly elongated southeastwards, extending to $\sim13$ kpc. The blue- and red-shifted velocities primarily appear in opposite directions, suggesting the presence of a rotating disk. However, the red-shifted velocities reach significantly higher values than their blue-shifted counterparts, indicating a structural asymmetry in the disk (e.g., due to external interactions; see Section \ref{sec:tidal}).\par
\begin{figure*}[t]
   \centering
   \resizebox{\hsize}{!}
        {\includegraphics[width=1.00\linewidth]{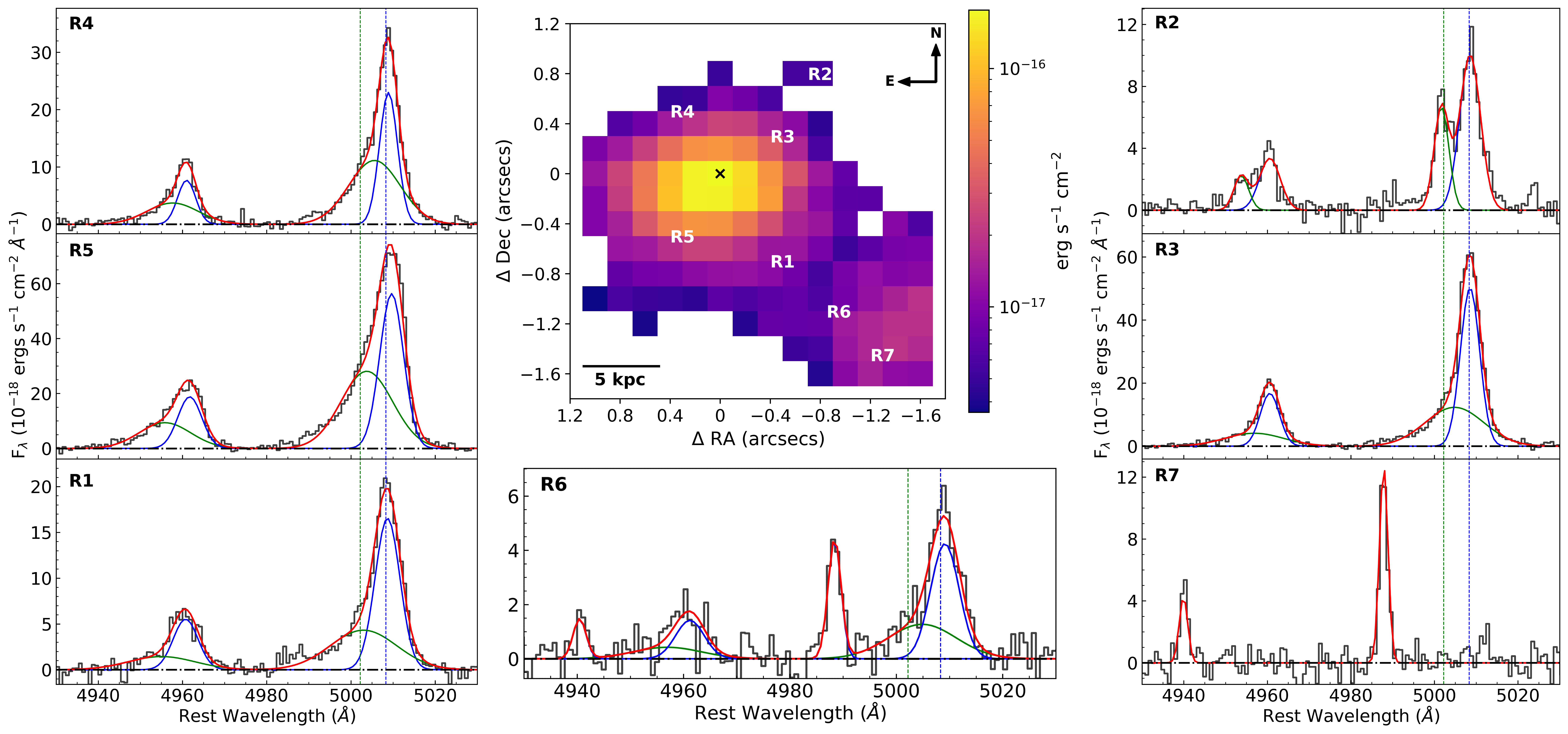}}
      \caption{Spatially resolved kinematics of the extended [\ion{O}{III}] emission. Center: Flux map generated by integrating the rest-frame spectrum between 4985 and 4995 Å. This specific wavelength range was chosen to isolate the high-velocity wing of the outflows and the highly-blueshifted narrow feature (see Section \ref{sec:tidal}). \textbf{Surrounding panels}: Continuum-subtracted spectra (black) and best-fit models (red) for the labeled regions. Spectra are extracted from $2 \times 2$ spaxel apertures. Unresolved emission from the BLR and Comp.\ C was subtracted along with the continuum to isolate the spatially resolved systemic (blue) and outflow (green) components. The dashed vertical lines correspond to the centroids of these components in the unresolved nuclear spectrum (Figure \ref{fig:central_model}).}
         \label{fig:specmaps}
   \end{figure*}

 
Component B (bottom row, Figure \ref{fig:maps}) shows high blueshifted centroid velocities (up to $-650$ km s$^{-1}$) that are generally accompanied by large widths (FWHM $>$ 600 km s$^{-1}$). These kinematic signatures are consistent with the outflowing component identified in the nuclear spectrum. The v$_{50}$ map in this case highlights two regions of particular interest, which we mark as B1 (black contour) and B2 (red contour). Region B1, which aligns with the brightest emission detected in panel (c) of Figure \ref{fig:psfsubmaps}, traces the bulk of the highly blueshifted outflowing gas with v$_{50}$ $\lesssim-$250 km s$^{-1}$. This region also corresponds to higher FWHM values ($\gtrsim$ 1000 km s$^{-1}$), and extends out to $\sim$ 8 kpc southwestwards. In this direction, both velocities and the width decrease slightly as we move away from the center, before increasing abruptly at the edge. As we discuss in Section \ref{sec:tidal} (and show in panels R6 and R7 of Figure \ref{fig:specmaps}), this behavior does not reflect a drastic change in the kinematics of the outflowing gas, but is rather due to contamination from unrelated external emission. \par
Region B2 displays blueshifted velocities similar to B1, but with smaller widths (FWHM $\sim$ 250 km s$^{-1}$). As shown in panels R1 and R2 of Figure \ref{fig:specmaps}, the continuum-subtracted spectra highlight a stark contrast in line widths, revealing a distinct transition from a fading broad wing in the central region to a new, narrow peak in B2. Because our fitting algorithm operates on a spaxel-by-spaxel basis, it translates this change in spectral shape into the visually abrupt boundary seen in the parameter maps (Figure \ref{fig:maps}). However, we also note that this feature lies at the edge of our field of view, where its spatial extent is comparable to the instrumental PSF. Thus, while the change in the line profile of the gas in region B2 is clear, its spatial morphology should be treated with caution.\par
One physical interpretation of the narrow profile in B2 is that it represents gas unrelated to the primary outflow. Instead, it may simply be ambient gas illuminated by the AGN. We note, however, that the centroid velocity shift of this gas is similar to that of the outflowing gas in B1, which could suggest a common physical origin for the two. This connection is also supported by the highly blueshifted channel map in panel (c) of Figure \ref{fig:psfsubmaps}, where the bright emission originates from B1 and the region connecting B2 to the central source. Under the assumption of a common physical origin, the narrower profile of B2 may reflect outflowing gas escaping through a less dense path within the ISM, where it retains high blueshifted velocities but exhibits reduced line widths due to laminar motion \citep[e.g., ][]{liu2013observations,zhao2023discovery}.\par
Overall, the spatially resolved kinematics of Comp.\ B highlight a kiloparsec-scale, wide-angle outflow dominated by region B1. This highly blueshifted emission directly traces the approaching face of an outflow expanding toward our line of sight. In this geometry, if a redshifted counterpart to this outflow structure exists, it would likely be obscured by the large-scale galactic disk traced by Comp.\ A, which is expected to be nearly face-on for a type-1 AGN like PKS 0352.\par

\begin{figure}[t]
   \centering
   \resizebox{\hsize}{!}
        {\includegraphics[width=0.80\linewidth]{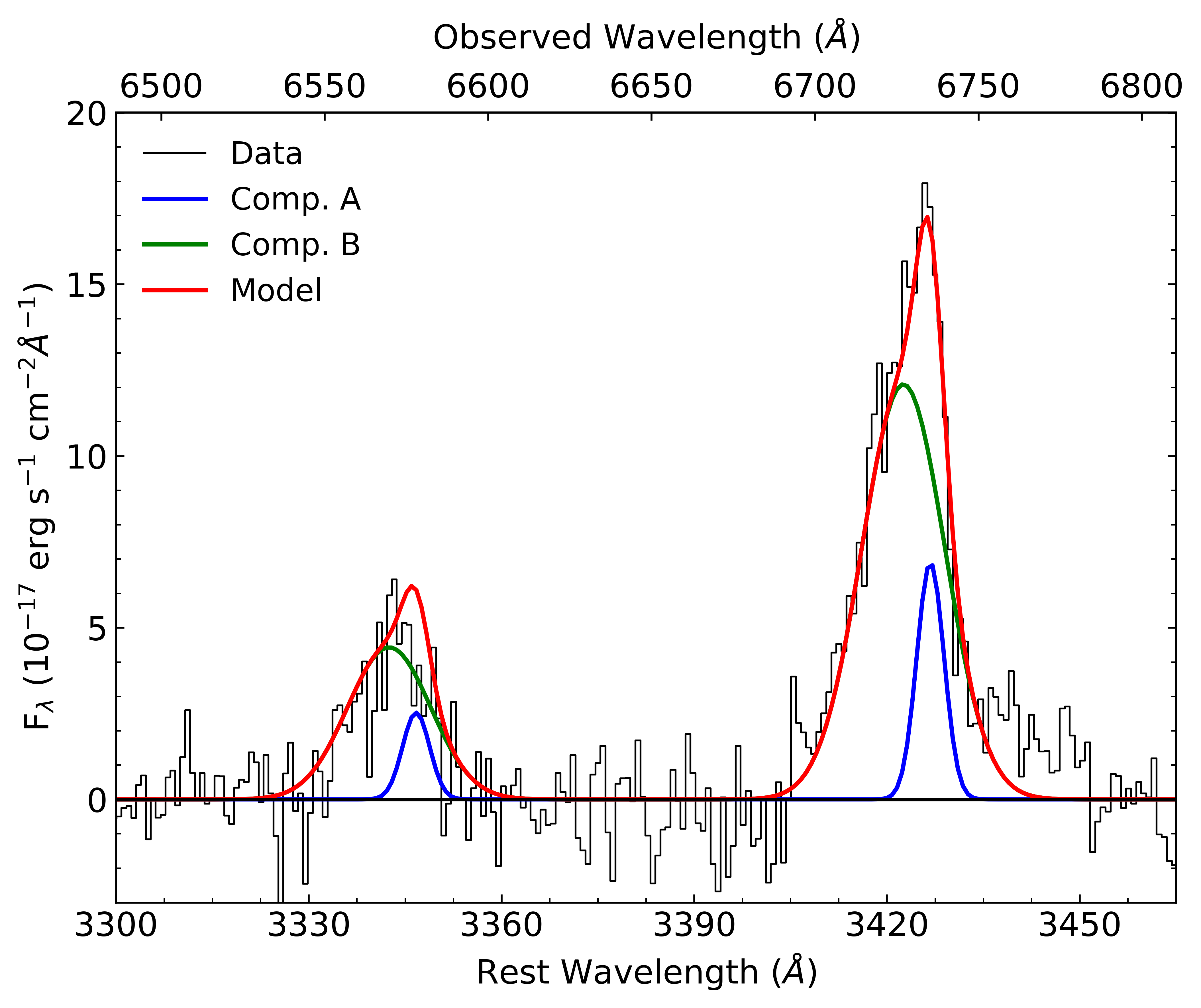}}
      \caption{Continuum subtracted SDSS spectrum of the [\ion{Ne}{V}] region in PKS 0352. The observed spectrum is shown in black and our best-fit model in red. The blue and green curves trace the systemic and outflowing components, respectively, and show similar kinematics to their [\ion{O}{III}] counterparts in Figure \ref{fig:central_model}.}
         \label{fig:NeV_model}
   \end{figure}

\subsection{Unresolved UV Emission} \label{subsec:UVnuclear}

The SDSS spectrum of PKS 0352 allows us to probe the unresolved central region in the rest-frame near-UV. In this spectral region, we identify prominent emission lines from other forbidden transitions, specifically the [\ion{Ne}{V}] $\lambda \lambda$ 3426, 3346 and [\ion{Ne}{III}] $\lambda \lambda$ 3869, 3967 doublets. The detection of the [\ion{Ne}{V}] doublet is of particular interest, as the production of Ne$^{4+}$ requires photons with energies $> 97.1$ eV, significantly higher than the 35.12 eV required for O$^{2+}$. Furthermore, the critical density of the [\ion{Ne}{V}] $\lambda$3426 line ($n_{\textrm{crit}}\approx~$1.5 $\times$ $10^7 \textrm{ cm}^{-3}$) is more than an order of magnitude higher than that of [\ion{O}{III}] $\lambda$5007 ($n_{\textrm{crit}}\approx~$7 $\times$ $10^5 \textrm{ cm}^{-3}$). The [\ion{Ne}{V}] emission thus serves as a tracer of the high ionization and high density phase of the NLR gas.\par
To accurately model these lines, we first subtract the underlying continuum using the  Bayesian AGN Decomposition Analysis for SDSS Spectra \citep[BADASS;][]{sexton2021bayesian} tool. We then fit the emission lines in the continuum-subtracted spectrum using multiple Gaussian components, fixing the flux ratios of the [\ion{Ne}{V}] and [\ion{Ne}{III}] doublets to their theoretical values of 2.7 and 3.1, respectively. Based on the Bayesian criteria described in Section \ref{subsec:nuclear}, we find that both these features are best reproduced with a two-component model. We report their derived kinematics in Table \ref{table:nuclear_kinematics} and show the best-fit model for the [\ion{Ne}{V}] doublet in Figure \ref{fig:NeV_model}. \par
Comparison of the [\ion{Ne}{V}] kinematics with the nuclear [\ion{O}{III}] emission reveals striking similarities. The host galaxy component (Comp.\ A) shows consistent velocity shift and width values in both cases. Similarly, the outflowing component (Comp.\ B) shows a velocity shift remarkably consistent with that of the [\ion{O}{III}] outflow, while being approximately 20\% broader. We note that the SDSS spectrum is derived from a 3-arcsecond diameter fiber, covering a significantly larger physical area than the \ang{; ; 0.6} $\times$ \ang{; ; 0.6} extraction region used for the central [\ion{O}{III}] emission. The close agreement between the overall kinematics, despite the aperture difference, strongly suggests that the bulk of the [\ion{Ne}{V}] emission comes from within the inner $\sim$ 2.4 kpc. This is consistent with spatially resolved studies of nearby AGN, where the coronal line region is found to be compact \citep[e.g. ][]{ramos2017infrared, rodriguez2020700}. More importantly, however, this shows that the high-ionization gas traced by \ion{Ne}{V} has a similar physical origin as the \ion{O}{III} gas and is tracing the inner, higher-ionization regions of the two distinct large-scale structures revealed in Figure \ref{fig:maps}. \par
The [\ion{Ne}{III}] emission also displays a kinematic profile broadly consistent with both [\ion{O}{III}] and [\ion{Ne}{V}], though with a possible trend towards smaller values for both the velocity shift and width. These deviations are expected, as unlike [\ion{Ne}{V}], the [\ion{Ne}{III}] emission need not be completely confined within the central region. If the [\ion{Ne}{III}] emission extends beyond \ang{; ; 0.6}, the SDSS spectra will include contribution from the slower, less turbulent gas at larger radii, thus weighing down the integrated values.
\begingroup
\setlength{\tabcolsep}{3pt}
\renewcommand{\arraystretch}{1.5}
\begin{table}
\caption{Properties of the outflowing components detected in PKS 0352}
\centering
\begin{tabular}{lccccc}
\hline
Comp.\ & $v_\textrm{max}$ & $R$ & $\dot{M}$  & log($\dot{P}$) & log($\dot{E}_{k}$)  \\ 
- & [km s$^{-1}$] & [pc] &[M$_{\odot}$ yr$^{-1}$] & [dyne] & [erg s$^{-1}$] \\
\hline
S2 & $-3700$ & $9^{+5}_{-5}$ & $7^{+9}_{-5}$ & $35.14^{+0.3}_{-0.5}$ & $43.33^{+0.3}_{-0.5}$\\
C & $-3900$ & $40-2400$ & 1--17 & 34.2--35.6 & 42.5--43.9\\
S1 & $-2100$ & $520^{+300}_{-150}$ & $10.6^{+21}_{-8}$ & $35.09^{+0.5}_{-0.5}$ & $43.06^{+0.5}_{-0.5}$\\
B$^a$ & $-1000$ & $8400$ & $10.1^{+40}_{-5}$ & $34.92^{+0.7}_{-0.3}$ & $42.73^{+0.7}_{-0.3}$\\
\hline
\end{tabular}
\tablecomments{The outflow components are arranged in order of increasing distance from the central source. Properties of the absorption outflows (S1 and S2) are taken from \citet{miller2020hst} and have been corrected to match the systemic redshift used in this work.\\ $^a$ The reported $v_{\textrm{max}}$ and $R$ values correspond to the farthest individual spaxel in which Comp.\ B is detected with a S/N $\geq$ 2. The actual maximum extent of the outflowing gas is larger than $R$, as confirmed by panel R6 of Figure \ref{fig:specmaps}, which is at a distance of $\sim$ 10.4 kpc. The uncertainties reported here stem from assumptions regarding $n_e$ (see text) and are significantly larger than the measurement uncertainties.}
\label{table:outflow_properties}
\end{table}
\endgroup

\section{Outflow Rates and Energetics} \label{sec:energetics}

Using the spatially and kinematically resolved properties of the outflowing gas, we can estimate its mass and energetics. The luminosity of the [\ion{O}{III}] $\lambda$5007 line directly traces the mass of the ionized gas responsible for the emission. This relationship takes the form \citep[e.g.,][]{carniani2015ionised, veilleux2020cool}:

\begin{equation} \label{eqn:mass}
    M_{\textrm{[O III]}} = 5.3~\times~10^7\textrm{ M}_{\odot} \left(\frac{L_{[\textrm{O III}]}}{10^{44} \textrm{erg s}^{-1}}\right) \left(\frac{n_e}{10^3 \textrm{cm}^{-3}}\right)^{-1} \left(\frac{C_e}{10^{[\textrm{O/H}]}}\right)
\end{equation}
where $L_{\textrm{[O III]}}$ is the luminosity of the [\ion{O}{III}] $\lambda$5007 emission line and $n_e$ is the electron number density of the ionized gas. $C_e = \langle n_e \rangle ^2/\langle n_e^2\rangle$ is the electron density clumping factor and is assumed to be 1. The term $10^{[\textrm{O/H}]}$ represents the oxygen-to-hydrogen abundance ratio relative to the solar value. Because \citet{miller2020hst} found that the photoionization models require super-solar metallicities for the absorption outflows in PKS 0352, we adopt their value of $10^{[\textrm{O/H}]} = 3$ for consistency.\par
As our KMOS observations lack density-sensitive diagnostic lines, $n_e$ represents a significant source of uncertainty. Typical values of $n_e$ reported by recent spatially-resolved measurements using the [\ion{S}{II}] $\lambda \lambda$ 6716, 6731 doublet are in the range of 100$-$1000 cm$^{-3}$ \citep[e.g.][]{veilleux2023first, vayner2023first, cresci2023bubbles}, although higher values have been reported for nearby AGN using other diagnostics \citep[e.g.,][]{davies2020ionized,revalski2022quantifying}. For Comp.\ B, we adopt a fiducial value of $n_e = 500 \textrm{ cm}^{-3}$, consistent with similar studies of large-scale outflows in luminous quasars \citep[e.g.,][]{carniani2015ionised, loiacono2019multiwavelength, belli2024star}. To account for the effect of this assumption on the measured outflow energetics, we include the systematic error due to the possible $n_e$ range in the values reported in Table \ref{table:outflow_properties}. Finally, we measure an integrated luminosity $L_{[\textrm{O III}]} = 6.3 \times 10^{43} \textrm{ erg s}^{-1}$ for Comp.\ B, which results in an ionized gas mass of $M_{\textrm{[O III]}}= 2.2~\times~10^{7} M_{\odot}$. \par
 
 The total outflow rate is calculated by summing over individual spaxels and can thus be written as:
\begin{equation}\label{eqn:outflowrate}
  \dot{M} = \sum_i \frac{M_i v_i}{R_i},
\end{equation}
where $M_i$, $v_i$ and $R_i$ are the gas mass, velocity, and distance from the quasar for a given spaxel $i$. Similarly, the momentum rate and kinetic luminosity are given by:
\begin{align}\label{eqn:kinlum}
\dot{P} &= \sum_i \dot{M_i} v_i &  \dot{E}_k &= \frac{1}{2} \sum_i \dot{M_i} v_i^2
\end{align}
Determining the intrinsic outflow velocity from the measured LOS kinematics requires detailed geometric modeling, which is not feasible given the spatial resolution of the current dataset. We thus follow the standard convention of adopting  $v_{\textrm{max}}$ = $v_{0}~-$ 2$\sigma$ as the outflow velocity. \par
These measurements are sensitive to PSF smearing, because $M_i$ in Equation (\ref{eqn:outflowrate}) must only account for the mass of the gas physically represented by the spaxel. We thus adopt a two-step approach to address this for Comp.\ B. First, for the unresolved nuclear emission (from the central $3\times3$ spaxels, Figure \ref{fig:central_model}), we derive the total outflowing flux by applying an aperture correction to the model flux based on our PSF template. Next, for the spatially resolved emission, we use the same PSF template to scale the model flux from the central spaxel and subtract it from the model fluxes for all spaxels where Comp.\ B is detected. This residual flux is then used to obtain the mass in each spaxel. The final values obtained by summing up the contribution from spatially unresolved and resolved emission for Comp.\ B are $\dot{M}$ = 10.1  M$_{\odot}$ yr$^{-1}$, $\dot{P}$ = 8.3 $\times$ $10^{34}$ dyne and $\dot{E}_k$ = 5.4 $\times$ 10$^{42}$ erg s$^{-1}$. \par
As reported in Section \ref{subsec:resolved}, the other outflowing component (Comp.\ C) remains spatially unresolved by our KMOS observations. This constrains its maximum extent to less than half the PSF FWHM, corresponding to $R_{\textrm{max}}\simeq2.4 \textrm{ kpc}$. Because the [\ion{O}{III}]$\lambda$5007 line arises from a forbidden transition, the bulk of its emission cannot originate from a region where $n_e$ exceeds the critical density, $n_{\textrm{crit}}\approx~$7 $\times$ $10^5 \textrm{ cm}^{-3}$. \citet{miller2020hst} reported a density of $n_e$ = $1.3\times10^{7}$ cm $^{-3}$ for the absorption outflow S2, located at a distance of 9 pc. Using this as a representative value and assuming a density profile of an isothermal sphere ($n_e\propto R^{-2}$), we estimate a minimum emitting radius for Comp.\ C of $R_{\textrm{min}}\simeq 40 \textrm{ pc}$. By combining these radial constraints for Comp.\ C with the corresponding values of $n_e$ ($7\times 10^5 \textrm{ cm}^{-3}$ at $R_{\textrm{min}}$ and $500 \textrm{ cm}^{-3}$ at $R_{\textrm{max}}$), we derive a plausible range for the outflow rates and energetics using Equations (\ref{eqn:outflowrate}) and (\ref{eqn:kinlum}). This yields $\dot{M}$ = $1\textrm{--}17~M_{\odot}$ yr$^{-1}$, $\dot{P}$ = $(2\textrm{--}42)\times$ $10^{34}$ dyne and $\dot{E}_k$ = $(4\textrm{--}82)\times$ 10$^{42}$ erg s$^{-1}$. \par
We note that for both outflow components, the derived energetics likely represent lower limits as they do not account for gas residing in other ionization states or phases. The existence of a more highly ionized phase in the inner regions is evident from the detection of the unresolved [\ion{Ne}{V}] emission that is kinematically associated with Comp.\ B. Similarly, at larger scales, the neutral/molecular phase of the outflow could contain a considerable fraction of the total gas mass \citep[e.g.,][]{rupke2013multiphase,fiore2017agn}. \par  
\section{Connection between the absorption and emission outflows} \label{sec:physcon}
Combining high-resolution absorption spectroscopy with spatially resolved emission line kinematics has revealed four distinct outflowing components in PKS 0352 that span a wide range in ionization (from \ion{Mg}{X} to \ion{O}{III}). In absorption, two high-ionization outflows are detected with velocities of $-3700 \textrm{ km s}^{-1}$ (S2) and $-2100 \textrm{ km s}^{-1}$ (S1), at distances $\sim$ 10 and 500 pc, respectively. In emission, we detect a fast, unresolved outflow with a velocity of $-3900 \textrm{ km s}^{-1}$ (Comp.\ C), along with a galactic-scale blueshifted outflow that extends to $\sim$ 10 kpc with a velocity of $\sim1000\textrm{ km s}^{-1}$. Despite a three order of magnitude range in spatial extent across these outflows, their kinematics and energetics show remarkable consistency, suggesting that they might be tracing the integrated history of connected feedback events from nuclear to galactic scales. \par
We first note the striking similarity between the maximum velocities of the nuclear absorption outflow S2 ($v_{\textrm{max}}\sim-$3700 km s$^{-1}$) and the unresolved [\ion{O}{III}] outflow traced by Comp.\ C ($v_{\textrm{max}}\sim-$3900 km s$^{-1}$). Absorption measures the true de-projected velocity of the outflowing gas along our LOS to the quasar. On the other hand, $v_{\textrm{max}}$ for a volume-integrated emission line measures the velocity of the most blueshifted fraction of the gas, projected along our LOS. As long as some portion of the emitting outflow lies along our LOS, either due to a wide-angle geometry or a favorable orientation, the emission $v_{\textrm{max}}$ will suffer from minimal projection effects and provide a similar kinematic measure as the absorption feature. For a Type-1 quasar, this geometric condition is likely to be met, particularly at the nuclear scales being probed \citep[e.g.,][]{elvis2000structure}. It is then unlikely that two unrelated outflow events would lead to such a remarkable match in their intrinsic velocities. Instead, this suggests that S2 and Comp.\ C may both represent the same phase in the kinematic evolution of the outflowing gas as it traverses the inner regions of the galaxy. This interpretation favors a compact origin for Comp.\ C, placing it closer to its minimum emitting radius of 40 pc rather than its maximum allowed extent of 2.4 kpc. \par
As discussed in Section \ref{sec:energetics}, between the locations of S2 and Comp.\ C, the density must drop by more than an order of magnitude, despite which the outflowing gas appears to maintain a nearly constant velocity ($v \sim -3800$ km s$^{-1}$). This behavior is physically expected if the density profile in the inner regions is relatively steep, approaching that of an isothermal sphere. Under these conditions, the mass accumulated by the outflow increases roughly linearly with radius ($M\propto R$), and thus the rate at which mass is swept up depends strictly on the outflow's velocity ($\dot{M}=M/(R/v) \propto v$). Depending on the driving mechanism, the AGN continuously injects a constant rate of either momentum ($\dot{P}_{\textrm{in}}$) or kinetic energy ($\dot{E}_{\textrm{in}}$) into the outflow \citep[e.g.,][]{king2011large,faucher2012physics}. As the rate of change in momentum and energy for the outflowing gas scale as  $\dot{P}=\dot{M}v\propto v^2$ and $\dot{E}_k=\frac{1}{2}\dot{M}v^2\propto v^3$, the constant injection of either $\dot{P}_{\textrm{in}}$ or $\dot{E}_{\textrm{in}}$ naturally leads to a constant velocity expansion, as the driving force continues to balance the ongoing mass-loading from the ISM. However, as this outflow reaches larger distances where the host galaxy's density profile begins to flatten, the amount of swept-up ISM increases at a faster rate. As the accumulated mass begins to dominate, the outflowing gas must decelerate as it plows through the galaxy. We interpret the absorption system S1, detected at $\sim$ 500 pc with a reduced velocity of $v_{\textrm{max}} \sim~-$2100 km s$^{-1}$, as an indication of this transitioned mass-loaded phase. \par
The final stage of this cycle is then likely revealed by Comp.\ B, which is detected as unresolved nuclear emission in [\ion{Ne}{V}] and as spatially resolved emission in [\ion{O}{III}], extending out to $\gtrsim$ 8 kpc. Beyond S1 (500 pc), the outflow would continue to load mass as it pushes through the dense inner ISM. Within the central 2.4 kpc, the outflowing gas shows clear signs of further deceleration across multiple ionization phases. We detect the higher-ionization gas in [\ion{Ne}{V}] at $v_{\text{max}} \sim -1500$ km s$^{-1}$, alongside the lower-ionization [\ion{O}{III}] emission at $v_{\text{max}} \sim -1300$ km s$^{-1}$. As the outflow expands beyond this central region, the [\ion{O}{III}] velocity decreases only slightly, maintaining $\sim-$1000 km s$^{-1}$ out to 8 kpc. This minimal deceleration over a distance of 6 kpc suggests that the gas may have broken out of the dense inner ISM and is traveling along the path of least resistance \citep[e.g., ][]{zubovas2014energy}. In the absence of a dense confining medium, the outflowing gas would expand laterally and form a wide-angle structure similar to that seen in Figure \ref{fig:maps}. Because the outflow would no longer be sweeping up significant mass in this scenario, its velocity drops only minimally as the gas continues to climb out of the host galaxy's gravitational potential. \par
This proposed connection, based on the kinematics of the outflowing gas across various spatial scales, is also supported by the measured momentum fluxes (Table \ref{table:outflow_properties}). These fluxes remain consistent even as the maximum velocity of the outflowing gas drops by nearly a factor of 4, from $\sim -3700 \textrm{ km s}^{-1}$ at 9 pc to $\sim-1000 \textrm{ km s}^{-1}$ at 8.4 kpc. The nuclear absorption (S2) traces recently launched outflowing gas with a dynamic timescale ($\tau \sim R/v$) of $\tau_{\textrm{S2}} \sim 2.4\times10^3 \textrm{ yr}$. In contrast, the extended galactic scale emission traced by Comp.\ B represents an integrated timescale of $\tau_{\textrm{B}} \lesssim 8.2\times10^6 \textrm{ yr}$. Any multiscale energetic comparison thus fundamentally relies on the assumption of a quasi-steady-state feedback cycle, where the time-averaged radiative output of the central engine remains constant across these timescales \citep[e.g.,][]{gonccalves2008detection, schmidt2017statistical,Worseck2021}. Compared to previous works on multiscale outflows, the detection of absorption outflow S1 provides an additional constraint for PKS 0352, as it traces the intermediate scale between the nuclear and galactic flows, with $\tau_{\textrm{S1}} \sim 2.4\times10^5 \textrm{ yr}$. While distinct, finely-tuned episodic outbursts could in principle reproduce the observed multiscale properties, the consistency of the momentum rates across these three epochs favors the simpler interpretation that these components represent the integrated record of a sustained, quasi-steady-state feedback cycle. \par
\begin{figure}[t]
   \centering
   \resizebox{\hsize}{!}
        {\includegraphics[width=1.0\linewidth]{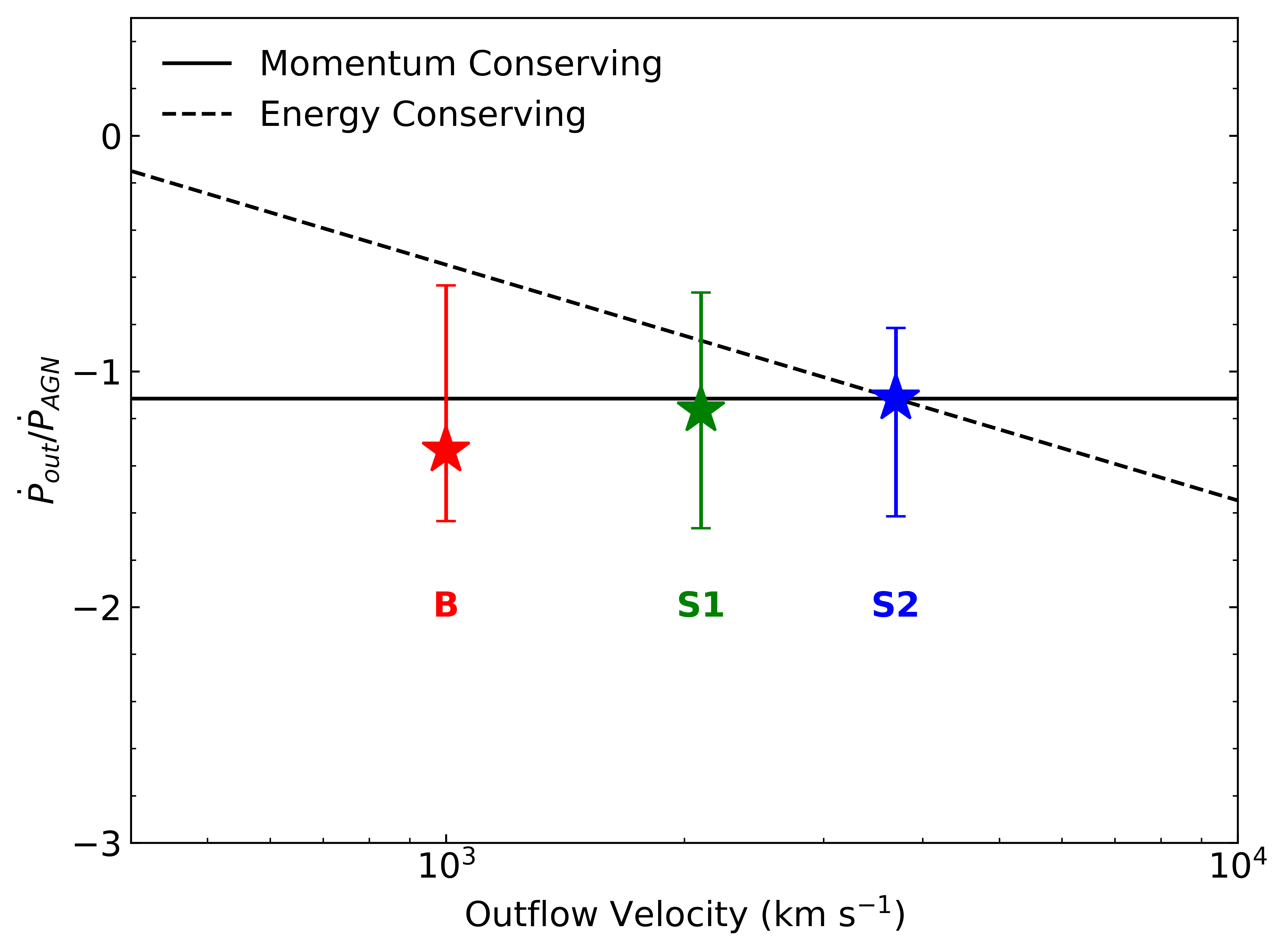}}
      \caption{Ratio of the outflow momentum fluxes ($\dot{P}_{\textrm{out}}$) and the radiative momentum flux of the quasar ($\dot{P}_{\textrm{AGN}} = L_{\textrm{bol}}/c = 1.8\times10^{36} \textrm{ dyne}$) as a function of outflow velocity. The data points show the outflow measurements across the three different spatial scales, along with the associated uncertainties. The solid and dashed black lines show the theoretical expectation in the case of momentum- and energy-conserving expansions, respectively. Their normalization is based on the innermost outflow S2. We omit Comp. C from this comparison as its velocity and allowed range for $\dot{P}_{\textrm{out}}/\dot{P}_{\textrm{AGN}}$ overlap with S2 (see Table \ref{table:outflow_properties}).}
         \label{fig:Pdot_compare}
   \end{figure}

In Figure \ref{fig:Pdot_compare}, we show the momentum boost ($\dot{P}_{\textrm{out}}/\dot{P}_{\textrm{AGN}}$) for the outflows at different spatial scales, as a function of their velocities. The solid black line represents a momentum-conserving expansion of the outflowing gas, whereas the dashed black line shows the expectation for an energy-conserving scenario, with $\dot{P} \propto v^{-1}$. Despite their close alignment with the momentum-conserving case, our measured $\dot{P}$ values cannot effectively rule out an energy-conserving flow due to their large uncertainties. In either case, our measurements reveal that $\dot{P}_{\textrm{out}}/\dot{P}_{\textrm{AGN}} \lesssim 0.1$ across all stages. Such low values for the momentum boost are not unusual for outflows in luminous quasars \citep[e.g.,][]{bischetti2019gentle, veilleux2023first} and suggest that in principle they can be driven solely by the radiation pressure of the quasar. If they are driven by relativistic nuclear winds instead, as predicted for energy-conserving models, the low momentum boost would suggest an inefficient coupling with the ISM, where the thermal energy of the driving wind vents through a porous medium rather than doing work on the bulk gas \citep[e.g.,][]{bourne2014black,torrey2020impact}.

\section{Discussion} \label{sec:discussion}

\subsection{Extreme [\ion{O}{III}] Kinematics in the Central Region} \label{subsec:extreme}
\begin{figure}[t]
   \centering
   \resizebox{\hsize}{!}
        {\includegraphics[width=1.0\linewidth]{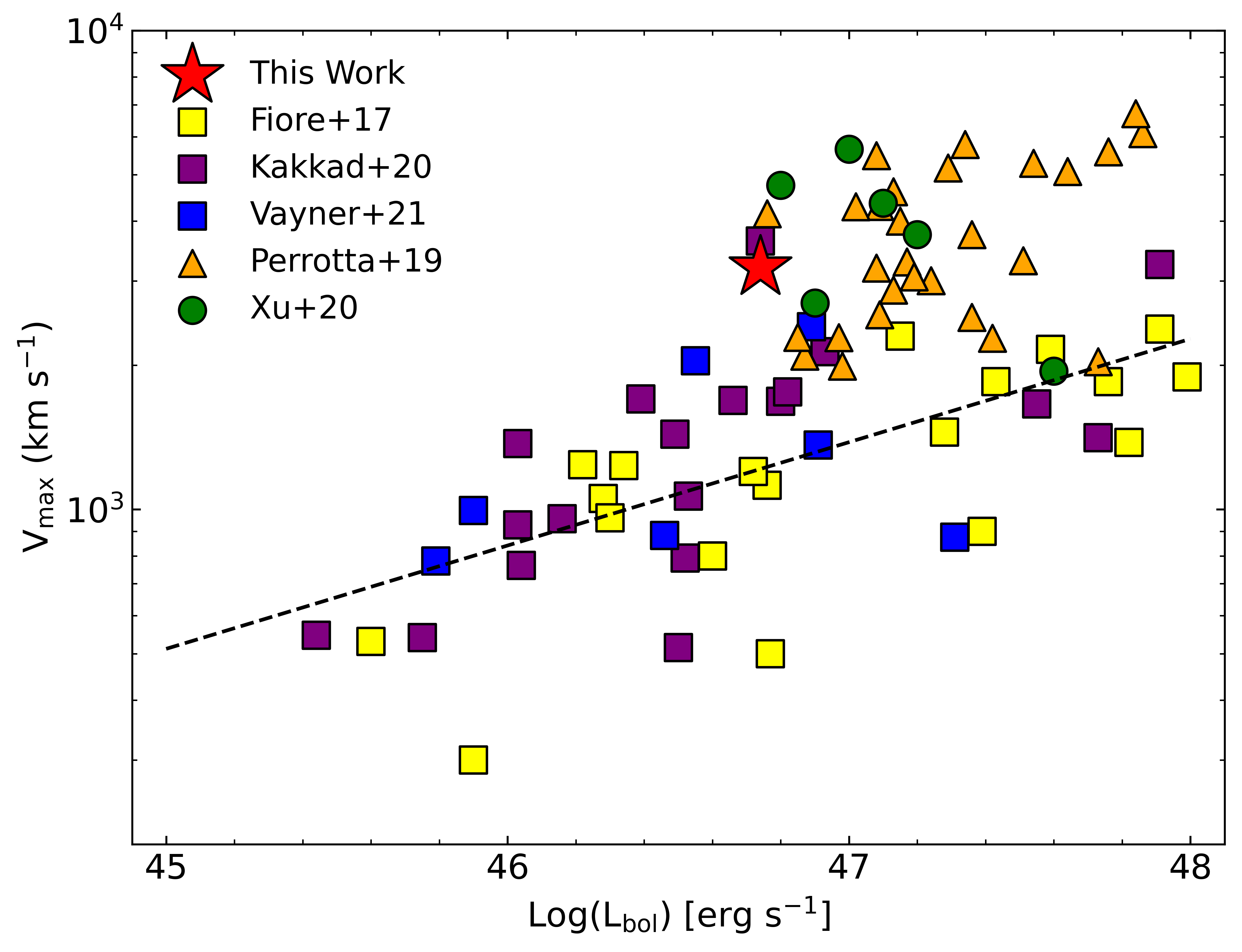}}
      \caption{Maximum [\ion{O}{III}] outflow velocity as a function of $L_{\textrm{bol}}$. PKS 0352 (red star) is plotted alongside samples of luminous, unobscured quasars (colored squares). These are compared against ERQs (orange triangles) and BALQSOs (green circles). The dashed line represents the best-fit correlation derived using only the colored squares ($L_{\text{bol}} \propto v_{\text{max}}^{4.6}$) and highlights the kinematic offset of the outflows in BALQSOs and ERQs from the typical quasar population at high luminosities.}
         \label{fig:vmax_compare}
   \end{figure}

As noted in Section \ref{subsec:nuclear}, a striking feature of the nuclear [\ion{O}{III}] emission is the extreme FWHM of Comp.\ C ($4130 \textrm{ km s}^{-1}$), which yields a maximum outflow velocity of $v_{\textrm{max}}=-3900 \textrm{ km s}^{-1}$. To facilitate comparison with the literature, we also derive a non-parametric maximum velocity for the integrated [\ion{O}{III}] outflow profile (Comp.\ B + Comp.\ C) using $v_{98}$, a metric that closely approximates $v_{\textrm{max}}$ for an individual Gaussian. It is defined as the velocity at the 98th percentile of the cumulative flux distribution (integrated from the redshifted to the blueshifted side), which we measure to be $v_{98}$ = $-3200 \textrm{ km s}^{-1}$. In Figure \ref{fig:vmax_compare}, we plot this alongside [\ion{O}{III}] outflows from various intermediate-redshift ($1 \lesssim z \lesssim 4$) quasar samples as a function of $L_{\textrm{Bol}}$. The outflow in PKS 0352 (marked with a red star) lies at the upper extreme of velocities observed in typical luminous, unobscured quasars \citep[from][marked with colored squares]{fiore2017agn, kakkad2020super, vayner2021spatially} \footnote{The \citet{fiore2017agn} sample includes sources compiled from \citet{Nesvadba2006,Nesvadba2008,Harrison2012,Genzel2014,carniani2015ionised,bischetti2017wissh}}. Notably, it aligns with a sample of BAL/mini-BAL quasars \citep[from][marked with green circles and hereafter simply referred to as BALQSOs]{xu2020evidence} \footnote{We exclude one object that shows positive absorption velocity in this sample, indicating either an infall or uncertainty in the systemic redshift.} which, similar to PKS 0352, exhibit high-velocity absorption outflows alongside their [\ion{O}{III}] emission. Furthermore, the maximum velocities measured in the BALQSO sample (now including PKS 0352) are similar to those seen in extremely red quasars \citep[ERQs, a highly obscured, dust-reddened population; orange triangles;][]{perrotta2019erqs}. The dashed line in Figure \ref{fig:vmax_compare} shows the best-fit correlation obtained using only the typical quasars (colored squares), which results in the relationship $L_{\textrm{Bol}}\propto v_{\textrm{max}}^{4.6\pm0.9}$, consistent with previous findings \citep[e.g.,][]{fiore2017agn}. Visually, both the BALQSO and ERQ samples exhibit a considerable offset from this relationship toward higher velocities. \par
We note that both these extreme samples exhibit complex, multi-component [\ion{O}{III}] outflows, and their maximum velocities are thus defined non-parametrically as $v_{\textrm{max}} = v_{98}$. In contrast, for the remaining objects, the maximum velocities are based on a single broad component representing the outflow, defined as $v_{\textrm{max}} = v_0 - 2\sigma \simeq v_{98}$. Because non-parametric measurements integrate over distinct spatial and kinematic components, they represent lower limits on the kinematics of the highest velocity fraction of the outflowing gas. Thus, the kinematic divergence at high luminosities between these samples and the typical quasars seen in Figure \ref{fig:vmax_compare} is likely an underestimate. \par
To statistically verify this divergence, we perform a two-sample 2D Kolmogorov–Smirnov test using the publicly available code \textsc{ndtest} \footnote{Written by Zhaozhou Li, \url{https://github.com/syrte/ndtest}.}. We test the BALQSO and ERQ samples individually against the base sample of luminous quasars and find $p-\textrm{values} < 10^{-4}$ in both cases, indicating less than a $0.01\%$ probability that they are drawn from the same parent distribution as the base sample. This departure strongly suggests that these samples are subject to conditions/environments that systematically lead to outflows with higher velocities compared to typical quasars. In the case of ERQs, \citet{perrotta2019erqs} show that the [\ion{O}{III}] kinematics strongly correlate with the red color of the host quasars, indicating that dust might play a crucial role in the large-scale acceleration of the outflows, likely via radiation pressure on dust grains \citep[e.g.,][]{murray2005maximum,costa2018quenching}. On the other hand, the BALQSO sample considered here is, on average, more than three magnitudes bluer than ERQs and thus dust is not expected to have as significant of an effect on their outflow kinematics. Instead, based on this work and \citet{xu2020evidence}, the [\ion{O}{III}] kinematics in the BALQSO sample appear to be correlated to the properties of their high-velocity absorption outflows. The [\ion{O}{III}] emitting region in these objects is estimated to be on a similarly compact scale ($\lesssim 500 \textrm{ pc}$) as the absorption outflow, further highlighting a physical connection between the two phases. \par
The extreme [\ion{O}{III}] velocities observed in the BALQSO sample could arise from orientation and projection effects. Under the assumption that all quasars have BAL outflows, the observed BALQSO population is a result of favorable orientation where the outflow solid angle intercepts our LOS \citep{elvis2000structure}. If the absorption and emission features trace the same physical outflow structure, as suggested for these objects by both \citet{xu2020evidence} and our current analysis, this geometry naturally selects for [\ion{O}{III}] outflows directed towards the observer, and thus their measured velocities will be close to the true de-projected outflow velocities. For non-BALQSOs, however, the outflows could be oriented at larger viewing angle relative to the LOS, which would lead to systematically lower velocity measurements due to geometric projection. The number of sources where this connection has been explored remains small, with PKS 0352 as the only object in this group with spatially resolved observations. Expanding this sample, particularly through high-resolution, spatially resolved observations, is essential to better understand the physical mechanism behind these high velocity emission outflows in BALQSOs.

\subsection{Evidence for External Interaction} \label{sec:tidal}
 
The spatially resolved analysis in Section \ref{subsec:resolved} reveals two kinematically distinct components within the central $\sim$ 10 kpc: the rotating disk of the host galaxy and the outflowing gas. Beyond this radius, extending south-west, we detect a third distinct feature. This feature is characterized by its significant blueshift ($v_{50} \sim -1200$ km s$^{-1}$) and narrow line width (FWHM $\sim$ 150 km s$^{-1}$).\par
In the middle panel of Figure \ref{fig:specmaps}, we present a narrow-band image constructed from the continuum-subtracted cube by integrating over the rest-frame wavelength range $4985-4995$ \r{A}. Within the central arcsecond, it captures the high-velocity wings of the broad outflowing component. However, at larger distances, it traces the distinct narrow feature. Region R1, which marks the detection edge of the spatially resolved [\ion{O}{III}] outflow in individual spaxels, also marks the transition zone where this narrow feature begins to appear on the blue wing of the outflowing component (around 4985 \r{A} in the rest frame). While the emission here is not significant enough to be detected as a separate feature, its contamination affects the derived kinematics of Comp.\ B as seen in Figure \ref{fig:maps}. To trace the spectral properties farther out in this direction, we extract integrated spectra from $2\times2$ apertures centered on regions R6 ($\sim 10.4$ kpc) and R7 ($\sim 16$ kpc). At R6, the highly blueshifted narrow feature becomes prominent. Due to the higher S/N of the integrated spectra, here we also detect faint emission from the host galaxy and the broad outflow, which are undetected in the individual spaxels in Figure \ref{fig:maps}. Moving to R7 near the edge of the field of view, this narrow feature continues to strengthen, while the emission associated with the quasar (both the host and the broad outflow) effectively disappears.\par
The low FWHM values for this component imply that this gas is dynamically cold, and therefore physically distinct from the high-velocity outflow. Furthermore, the high offset velocity is inconsistent with the systemic rotation of the host galaxy. The origin of this emission thus appears to be external. It could be associated with a merging companion galaxy, the presence of which is observed in 30–80\% of luminous quasars \citep{decarli2017rapidly,bischetti2021wissh}. Within our observational sensitivity, we do not detect underlying continuum emission in this region, thus ruling out a bright companion. We propose that this feature may represent a tidal tail formed by material stripped from a companion galaxy located outside our field of view, or tidal debris from a past fly-by or merger interaction. Such features, in different forms, have been observed in both nearby Seyferts \citep[e.g. Stephan's Quintet:][]{sulentic2001multiwavelength} as well as in luminous high-redshift quasars \citep[e.g. SDSSJ1652 + 1728:][]{wylezalek2022first, vayner2024first}. Alternatively, if the companion is relatively small, the feature could arise from the merging galaxy itself. Future observations, such as deep high-resolution imaging with HST, will be helpful in distinguishing between the different scenarios.

\section{Summary} \label{sec:summary}
In this work, we conduct a multiwavelength study of the ionized outflowing gas in the luminous $z = 0.9655$ quasar PKS 0352. Combining our VLT/KMOS IFS observations with previous HST/COS absorption analysis and SDSS observations allows us to trace the outflowing gas across three orders of magnitude in spatial scale. This multiscale approach yields the following main results:
\begin{enumerate}
    \item The [\ion{O}{III}] emission reveals two remarkable large-scale structures: The rotating disk of the host galaxy, extended to $\sim$ 13 kpc, and a wide-angle blueshifted outflow ($v_{\textrm{max}}\simeq -1000 \textrm{ km s}^{-1}$), extending beyond 8 kpc. Within the unresolved nuclear region, we also detect an extreme outflow component with $v_{\textrm{max}}= -3900 \textrm{ km s}^{-1}$, which is among the highest seen in luminous unobscured quasars.
    \item In the SDSS spectrum, we detect prominent [\ion{Ne}{V}] emission with a kinematic profile similar to that of the central [\ion{O}{III}] emission. This confirms that the high-ionization \ion{Ne}{V} gas is tracing the inner regions of the two large-scale structures revealed by the lower-ionization \ion{O}{III} gas.
    \item Linking the extended (Comp.\ B) and nuclear (Comp.\ C) emission outflows with the previously analyzed high-ionization absorption winds (S1 and S2) provides a cohesive evolutionary picture of the outflow. The kinematics support a scenario where the inner, constant-velocity expansion of the wind ($v_{\textrm{max}} \sim -3800$ km s$^{-1}$) is traced by both the absorption system S2 (at $\sim 9$ pc) and the unresolved emission wind (Comp.\ C at $\gtrsim 40$ pc). As the wind sweeps up the dense ISM, mass-loading forces it to decelerate to $-2100$ km s$^{-1}$ when it reaches $\sim$ 500 pc (S1). The outflow ultimately breaks out of the host galaxy's dense inner regions and travels along the path of least resistance, forming a wide-angle structure (Comp.\ B), as it maintains a velocity of $\sim -1000$ km s$^{-1}$ out to 8.4 kpc.
    \item Despite the significant deceleration of the outflowing gas across different spatial scales, the momentum fluxes remain remarkably consistent, within uncertainties, from parsec to kiloparsec scales. This consistency strongly implies that these distinct, multiscale outflow components represent the integrated history of a connected, sustained feedback cycle.
    \item We show that in BALQSOs where absorption and emission outflows appear to trace the same physical structure, [\ion{O}{III}] outflows are systematically faster than in typical quasars. This could result from the favorable line-of-sight orientation required to observe BAL troughs, which would also minimize geometric projection effects in the measured [\ion{O}{III}] velocities. Confirming this effect requires a larger sample of high-resolution, spatially resolved observations.
\end{enumerate}

Ultimately, PKS 0352 serves as an important first example in which the physical evolution of the outflowing gas is observed from nuclear to galactic scales through the complementary strengths of absorption and emission tracers. While our results establish a continuous kinematic link within the ionized gas, future observations targeting the molecular and neutral components will be valuable for completing the multiphase picture of this multiscale outflow. 

\begin{acknowledgements}
\section*{Acknowledgments}
We thank Sylvain Veilleux and Shobita Satyapal for helpful discussions. We also thank the anonymous referee for their constructive comments that improved the clarity of this paper. MS and NA acknowledge support from NSF grant AST 2106249, as well as NASA STScI grants AR-16600, AR-16601 and AR-17556. MB acknowledges financial support from PRIN MUR 2022 2022TKPB2P – BIG-z. This work is based on observations collected at the European Southern Observatory under program 0115.B-2246(A). The reduced dataset was obtained from the \cite{https://doi.org/10.18727/archive/38} Science Archive Facility. Funding for the Sloan Digital Sky Survey IV has been provided by the Alfred P. Sloan Foundation, the U.S. Department of Energy Office of  Science, and the Participating Institutions.
\end{acknowledgements}

\appendix

\begin{figure}[H]
   \centering
\includegraphics[width=0.50\linewidth]{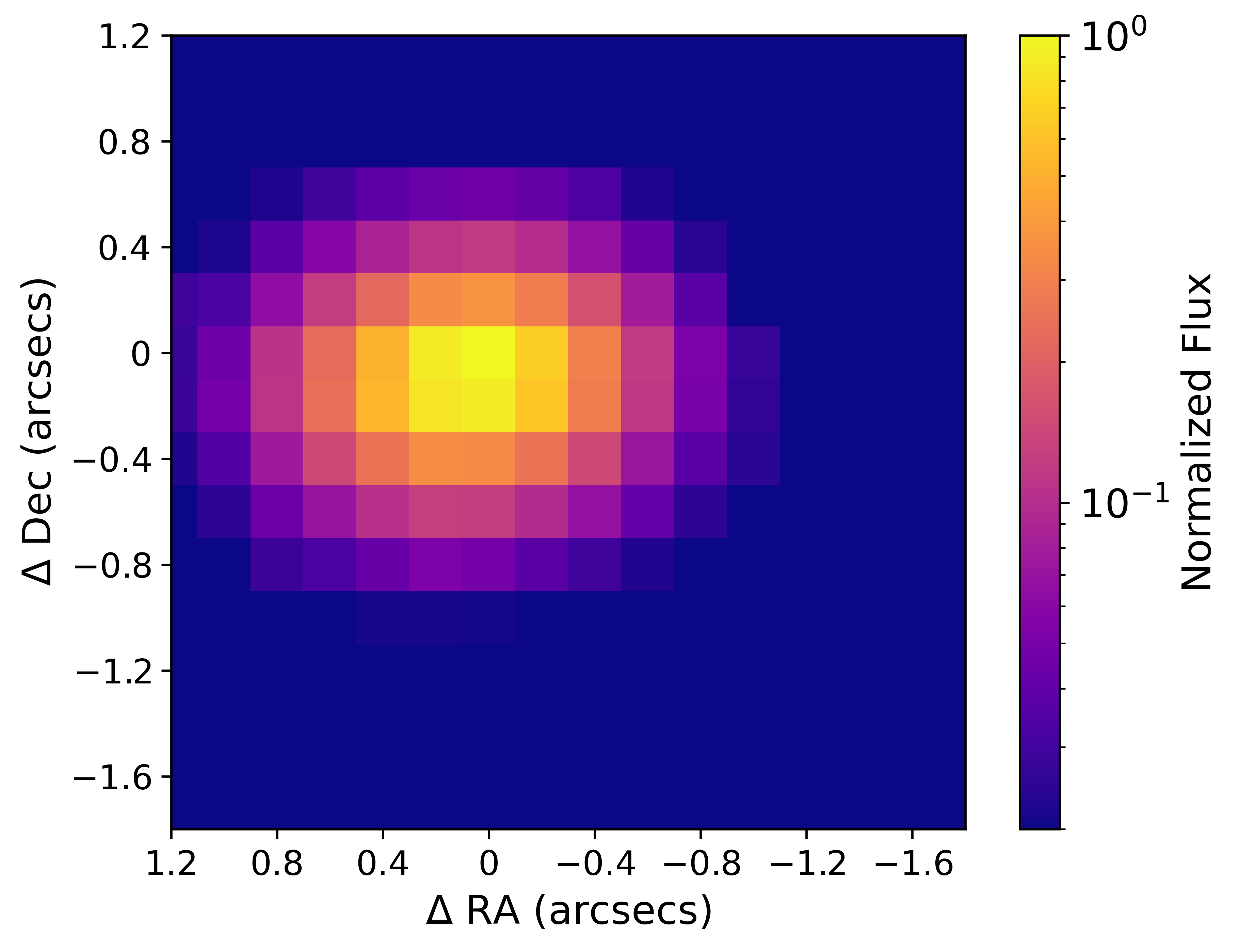}
    \caption{The empirical PSF template for PKS 0352 obtained using the line-free continuum region between 5070-5150 \r{A}. We apply a circular fit to estimate a FWHM of $\sim$ \ang{; ; 0.65}, but use the empirical template directly for removing the quasar point source.}
         \label{fig:psf_model}
   \end{figure}

\bibliography{PKS0352}{}
\bibliographystyle{aasjournal}

\end{document}